\documentclass[aps,prd,reprint,amsmath,amssymb,superscriptaddress,floatfix]{revtex4}
\usepackage{graphicx}
\usepackage{amsfonts}
\usepackage{amssymb}
\usepackage{rotating}
\usepackage{booktabs}
\usepackage{xcolor}
\usepackage{soul}
\usepackage{color}
\usepackage{slashed}
\usepackage{multirow}
\usepackage{makecell}
\usepackage{epsf}
\usepackage{ulem}
\usepackage{cancel}
\usepackage{color,bm}
\usepackage[colorinlistoftodos]{todonotes}

\usepackage[colorlinks=true,citecolor=cyan,urlcolor=blue,bookmarks=true,bookmarks=true,bookmarksopen=true,bookmarksnumbered=true,bookmarksopenlevel=3]{hyperref}

\definecolor{airforceblue}{rgb}{0.36, 0.54, 0.66}
\definecolor{steelblue}{rgb}{0.27, 0.51, 0.71}
\definecolor{amber}{rgb}{1.0, 0.49, 0.0}

\begin{document}

\title{ Double spin asymmetry in dihadron production in SIDIS off a longitudinally polarized nucleon target }
\author{Xuan Luo}
\author{Hao Sun\footnote{Corresponding author: haosun@dlut.edu.cn \hspace{0.2cm} haosun@mail.ustc.edu.cn }}
\author{Jing Li}
\author{Yi-Ling Xie}
\author{Tichouk}
\affiliation{ Institute of Theoretical Physics, School of Physics, Dalian University of Technology, \\ No.2 Linggong Road, Dalian, Liaoning, 116024, P.R.China }
\date{\today}

\begin{abstract}

In this paper we study the double longitudinal spin asymmetry of dihadron production in semi-inclusive deep inelastic scattering. 
We calculate an unknown twist- dihadron fragmentation function $\widetilde{D}^\sphericalangle$ within a spectator model 
which has been used successfully in describing the dihadron production in both the unpolarized and the single polarized processes. 
The collinear picture in which the transverse momentum of the final state hadron pair is integrated out has been considered. 
The $\cos\phi_R$ azimuthal asymmetry arising from the coupling $e_L H_1^\sphericalangle$ and the coupling $g_1 \widetilde{D}^\sphericalangle$ is studied. 
In general, $e_L(x)$ and its respective contribution vanish model-independently and $g_1 \widetilde{D}^\sphericalangle$ dominates in this asymmetry. 
Moreover, we have adopted the same choice as before for the quark mass m fixed to zero GeV, which eliminates the ${\rm Re}[F^{s*}F^p]$ contribution 
to the $\tilde{D}^\sphericalangle$. Using the model results obtained at the low hadronic scale and neglecting the QCD evolution effects we estimate the $\cos\phi_R$ 
asymmetry and compare with the COMPASS preliminary data. An estimate for the future electron ion collider is also presented.

\vspace{0.5cm}
\end{abstract}
\maketitle
\setcounter{footnote}{0}

\section{INTRODUCTION}
\label{I}

Azimuthal asymmetries in SIDIS are key observables to investigate the spin dependent substructure of the nucleon. 
Azimuthal asymmetries can be theorically connected to combinations of parton distribution functions (PDFs) and fragmentation functions. 
In recent years, it is of great interest to study the hadron pair production in semi-
inclusive deep inelastic scattering (SIDIS).
This process is encoded by the so-called dihadron fragmentation functions (DIFFs) describing the probability that a quark is fragmented into two hadrons.

The DiFFs were proposed for the first time in Ref.\cite{Konishi:1978yx} and their evolution equations have been studied in Refs.\cite{Vendramin:1981gi,Vendramin:1981te}.
The DiFF evolution equations were also studied in Ref.\cite{Ceccopieri:2007ip}, the first place where they were discussed as functions of the hadron pair invariant mass $M_h$.
Reference \cite{Collins:1994ax} introduced the transversely polarized DiFF as spin analyzer of the transversely polarized fragmenting quark, 
which later lead to the definition of $H_1^\sphericalangle$.
The basis of all possible DiFFs for two unpolarized detected hadrons has been given in Ref.\cite{Bianconi:1999cd}. 
The method of partial-wave analysis was introduced in Ref.\cite{Radici:2001na}, which is important to make the connection between the two produced hadrons clear.
Reference \cite{Bacchetta:2003vn} extended the analysis to subleading twist with transverse momentum being integrated out.Such analysis can be directly promoted into the transverse momentum-dependent case.
DiFFs have been used for investigating the transverse spin phenomena of the nucleon, 
and they can act as analyzers of the spin of the fragmenting quark \cite{Collins:1993kq,Jaffe:1997hf}.
Physicists started to focus on these functions by searching for a mechanism to extract the chiral-odd transversity distribution 
in an alternative and technically simpler way than the Collins effect \cite{Collins:1992kk}.
In this mechanism, the chiral-odd DiFF $H_1^\sphericalangle$ \cite{Radici:2001na,Bacchetta:2002ux} 
plays an important role in accessing transversity distribution $h_1$, as it couples with $h_1$ at the leading-twist level in the collinear factorization. 
In Refs.\cite{Bacchetta:2011ip,Bacchetta:2012ty,Radici:2015mwa,Radici:2016lam,Radici:2018iag}, the authors extracted $h_1$ from SIDIS and proton-proton collision data.
The first extraction of interference fragmentation functions was given by Ref.\cite{Courtoy:2012ry} in $e^+ e^-$ annihilation.
Meanwhile, the model predictions of the DiFFs by using the spectator model \cite{Bianconi:1999uc,Bacchetta:2006un,Bacchetta:2008wb} 
and by using the Nambu-Jona-Lasino quark model \cite{Matevosyan:2013aka,Matevosyan:2013eia,Matevosyan:2017alv,Matevosyan:2017uls} were performed.

Experimentally, azimuthal asymmetries of hadron pair productions with both the unpolarized target 
and the transversely polarized target have been measured by the HERMES Collaboration \cite{Airapetian:2008sk} and COMPASS Collaboration \cite{Adolph:2012nw,Adolph:2014fjw}.
Very recently, preliminary results on the azimuthal asymmetries in dihadron production were obtained by the COMPASS Collaboration \cite{Sirtl:2017rhi}.
Considering the dihadron cross section in a collinear factorization approach with the incident lepton beam being unpolarized or longitudinally polarized, 
only two modulation, the $\sin\phi_R$ and $\cos\phi_R$ angle dependence, remain. Here $\phi_R$ is the angle between the lepton plane and two-hadron plane. 
The $\sin\phi_R$ asymmetry was studied in a very recent paper \cite{Yang:2019aan} using a spectator model. Notice the $\cos\phi_R$ asymmetry was also studied in Ref.\cite{Yang:2019kqw}. 
This COMPASS measurement showed a clearly negative $\cos\phi_R$ asymmetry within experiment precision which was different from an obtained clearly positive $\sin\phi_R$ asymmetry.
In the parton model, in principle two source contribute to the $\cos\phi_R$ asymmetry, one is the coupling of the twist-three distribution $e_L$ 
and the twist-two DiFF $H_1^\sphericalangle$, the other is the twist-three DiFF $\widetilde{D}^\sphericalangle$ combined with the helicity distribution $g_1$.
The former contribution vanishes model independently: $e_L(x) = \int d^2 \vec{p}_T e_L(x,p_T) = 0$  \cite{Bacchetta:2006tn} since the T-odd PDF is forbidden by time-reversal invariance.
This is because after $p_T$ integration the Wilson line in the nonlocal quark bilinear expression, which defines the PDF, connects the quark fields along a straight line on the light cone. For a T-odd PDF this immediately implies that it vanishes.

In this paper, we use the spectator model results of the distribution functions and DiFFs to investigate the $\cos\phi_R$ asymmetry.
The only possible contribution of this asymmetry comes from $g_1 \widetilde{D}^\sphericalangle$. 
Thus we focus on the role of twist-three DiFF $\widetilde{D}^\sphericalangle$ encoding the quark-gluon-quark correlation.
It is reported in Refs.\cite{Kanazawa:2013uia,Metz:2012ct} that the contribution of the twist-three fragmentation function to the single spin asymmetry also plays an important role in SIDIS and proton-proton collision within the framework of the collinear factorization.
A phenomenological study \cite{Kanazawa:2014dca} presents that it is possible to simultaneously describe azimuthal asymmetries data in SIDIS and proton proton collisions \cite{Adams:2003fx,Abelev:2008af,Adamczyk:2012xd,Lee:2007zzh} by using collinear twist-three factorization. 
We adopt the spectator model \cite{Bacchetta:2006un} to calculate $\widetilde{D}^\sphericalangle$. 
In the calculation we consider the effect of the gluon rescattering at one loop level needed for nonzero twist-three quark-gluon-quark correlator for fragmentation as the authors have done in Refs.\cite{Lu:2015wja,Yang:2016mxl}. 
Applying the spectator model results for the distributions and DiFFs, we compute the $\cos\phi_R$ asymmetry and compare with the COMPASS preliminary data.

The paper is organized as follows. In Sec.\ref{II} we review the theoretical framework of the $\cos\phi_R$ azimuthal asymmetry of dihadron production 
in longitudinally polarized lepton beam scattered off a longitudinally polarized proton target. 
We apply the spectator model to calculate the twist-three DiFF $\widetilde{D}^\sphericalangle$ in Sec.\ref{III}. 
In Sec.\ref{IV}, we give the numerical results of the $\cos\phi_R$ azimuthal asymmetry at the kinematics of COMPASS as well as EIC. 
We make the summary for our work in Sec.\ref{V}.

\section{The $\cos\phi_R$ asymmetry of dihadron production in SIDIS}
\label{II}

\begin{figure}[htp]
\centering
\includegraphics[height=6.6cm,width=10.4cm]{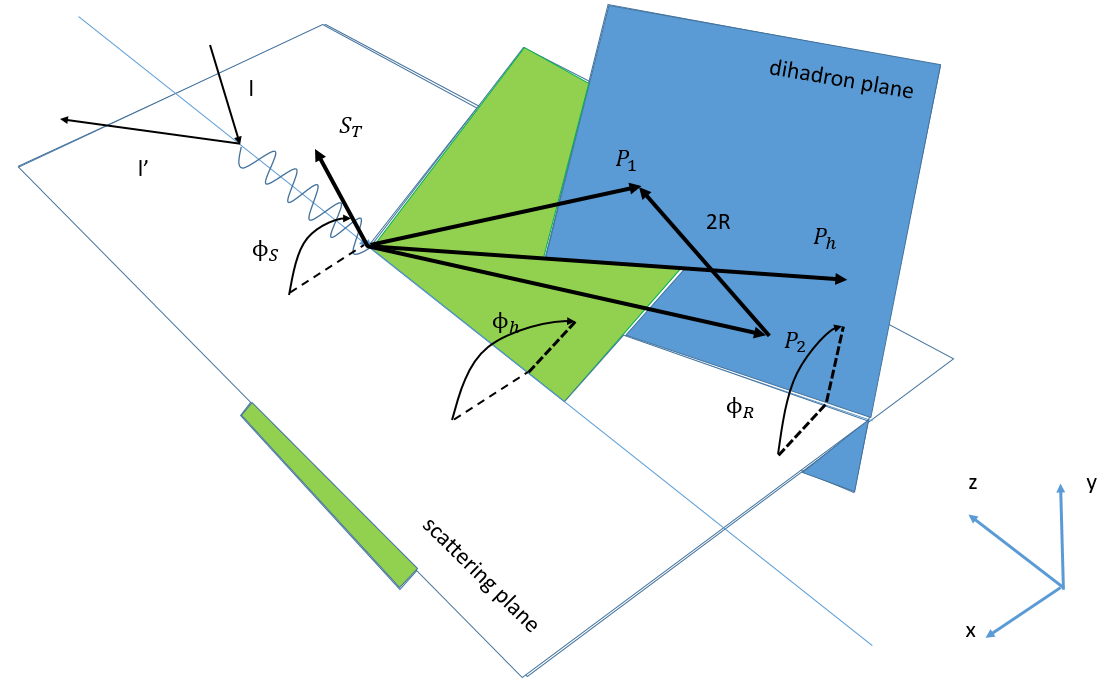}
\caption{Angle definitions involved in the measurement of the double longitudinal spin asymmetry in SIDIS production of two hadrons.}
\label{fig:1}
\end{figure}
As shown in Fig.\ref{fig:1}, we consider the dihadron SIDIS production
\begin{eqnarray}
\mu^\rightarrow(\ell)+p^\rightarrow(P) \to \mu(\ell')+h^+(P_1)+h^-(P_2)+X,
\end{eqnarray}
where a longitudinally polarized muon with momentum $\ell$ scatters off a longitudinally polarized target nucleon with mass $M$, polarization $S$ and momentum $P$, 
via the exchange of a virtual photon with momentum $q=\ell-\ell'$. Inside the target, the photon hits the active quark with momentum $p$ 
and the final state quark with momentum $k=p+q$ and then fragments into two leading unpolarized hadrons with mass $M_1, M_2$, and momenta $P_1, P_2$. 
To present the differential cross section with respect to dihadron-dependent structure function, we define the following kinematic invariants:
\begin{eqnarray}
\begin{aligned}
x   &=\frac{p^+}{P^+} \qquad  y=\frac{P \cdot q}{P \cdot \ell} \qquad z=\frac{P_h^-}{k^-}=z_1+z_2 \\
z_i &=\frac{P_i^-}{k^-} \qquad  Q^2=-q^2  \qquad s=(P+\ell)^2 \\
P_h &=P_1+P_2 \qquad  R=\frac{P_1-P_2}{2} \qquad  M_h^2=P_h^2.
\end{aligned}
\end{eqnarray}
Here, we describe the 4-vector in the light-cone coordinates as $a^\mu=(a^+,a^-,\vec{a}_T)$, 
where $\displaystyle a^\pm = \frac{a^0 \pm a^3}{\sqrt{2}}$ and $\vec{a}_T$ is the transverse component of the vector. 
Thus $x$ represents the light-cone fraction of target momentum carried by the initial quark, 
$z_i$ denotes the light-cone fraction of the fragmented quark carried by the hadron $h_i$. 
The light-cone fraction of fragmenting quark momentum carried by the final hadron pair is defined by $z$. 
Moreover, the invariant mass, the total momentum and the relative momentum of the hadron pair are denoted by $M_h$, $P_h$ and $R$, respectively.

It is convenient to choose the $\hat{z}$ axis according to the condition $\vec{P}_{hT}=0$. 
In this case, the momenta $P_h^\mu$, $k^\mu$ and $R^\mu$ can be written as \cite{Bacchetta:2006un}
\begin{eqnarray} \label{Eq:1} \nonumber
P_h^{\mu}&=&\left[P^-_h,\frac{M_h^2}{2P^-_h}, \vec{0}_T \right] \\ \nonumber
k^{\mu}  &=&\left[\frac{P_h^-}{z},\frac{z(k^2+\vec{k}_T^2)}{2P_h^-},\vec{k}_T \right] \\ \nonumber
R^{\mu}  &=&\left[-\frac{|\vec{R}|P^-_h}{M_h}\cos\theta,\frac{|\vec{R}|M_h}{2P^-_h}\cos\theta,|\vec{R}|\sin\theta\cos\phi_R,|\vec{R}|\sin\theta\sin\phi_R \right] \\
&=&\left[-\frac{|\vec{R}|P^-_h}{M_h}\cos\theta,\frac{|\vec{R}|M_h}{2P^-_h}\cos\theta,  \vec{R}_T^x, \vec{R}_T^y \right],
\end{eqnarray}
where
\begin{eqnarray}\label{Eq:2}
|\vec{R}| = \sqrt{\frac{M_h^2}{4}-m_\pi^2}.
\end{eqnarray}
Here $\phi_R$ is the angle between the lepton plane and the two-hadron plane and $m_\pi$ is the mass of pion. 
It is desired to notice that in order to perform partial-wave expansion, we have reformulated the kinematics in the CM frame of the dihadron system. 
$\theta$ is the cm polar angle of the pair with respect to the direction of $P_h$ in the target rest frame\cite{Bacchetta:2002ux}. 
Here we can find some useful relations as
\begin{eqnarray} \label{Eq:3}
P_h \cdot R &=& 0 \\
P_h \cdot k &=& \frac{M_h^2}{2z}+z\frac{k^2+k_T^2}{2} \\
R \cdot k &=& \left( \frac{M_h}{2z}-z\frac{k^2+k_T^2}{2M_h} \right) |\vec{R}| \cos\theta - k_T \cdot R_T \ \ .
\end{eqnarray}

We will consider the SIDIS process of longitudinally polarized muons off longitudinally polarized nucleon target. 
After integrating out the transverse momentum of the dihadron, the differential cross section for this process reads
\begin{eqnarray}\label{Eq:4}
\frac{d^6 \sigma_{\rm UU}}{d\cos\theta dM_h^2 d\phi_R dz dx dy} = \frac{\alpha^2}{Q^2 y} \left( 1-y+\frac{y^2}{2} \right) \sum_a e_a^2 f_1^a(x) D_1^a(z,M_{h}^2,\cos\theta)
\end{eqnarray}
and
\begin{eqnarray}\label{Eq:5} \nonumber
\frac{d^6 \sigma_{\rm LL}}{d\cos\theta dM_h^2 d\phi_R dz dx dy} &=&\frac{\alpha^2 S_L}{Q^2 y} 
\sum_a e_a^2 \bigg[ y(\frac{y}{2}-1)g_1^a(x)D_1^a(z,M_h^2,\cos\theta) + \left( 2y\sqrt{1-y} \right) \cos\phi_R \sin\theta \frac{|\vec{R}|}{Q} \\
& &
\left( \frac{M}{M_h} x e_L^a(x) H_1^{\sphericalangle,a}(z, M_h^2, \cos\theta) - \frac{1}{z} g_1^a(x)\widetilde{D}^{\sphericalangle,a}(z, M_h^2, \cos\theta) \right) \bigg],
\end{eqnarray}
where $S_L$ is the longitudinal spin component.
For convenience, we have indicated the unpolarized or longitudinally polarized states of the beam or the target with the labels U and L, respectively. 
In Eq.(\ref{Eq:4}), $f_1^a(x)$ and $D_1^a(z, M_h^2,\cos\theta)$ are the unpolarized PDF and unpolarized DiFF with flavor $a$. 
We have removed the vanishing contribution coming from $e_L^a(x) H_1^{\sphericalangle, a}$ in Eq.(\ref{Eq:5}), 
where $g_1^a(x)$ is the helicity distribution combined with the twist-three DiFF $\widetilde{D}^{\sphericalangle,a}$. 

The collinear DiFFs $D_1^a$ is extracted from the integrated quark-quark correlator $\Delta(z, R)$
\begin{eqnarray} \label{Eq:6} \nonumber
\Delta (z, R) \displaystyle &=&z^2 \sum \kern -1.3 em \int_X \;
\int \frac{d \xi^+}{2\pi} \; e^{\text{i} k\cdot\xi}\;
\langle 0|\, U^+_{[0,\xi]} \, \psi(\xi) \,|P_h,R; X\rangle
\langle X; P_h,R|\, \overline{\psi}(0)\,|0\rangle
\Big|_{\xi^- = \vec{\xi}_T  =0} \; \\
&=&\frac{1}{16\pi}\left\{ D_1 \slashed{n}_-+\cdots \right\},
\end{eqnarray}
where $\psi$ is the quark field operator and $U_{[b,c]}^a$ is the Wilson line running from $b$ to $c$ along $a$ to ensure the gauge invariance of the operator. 
$n_-$ denotes the negative light-like vector $n_-=[0,1,{\vec 0}_T]$. The twist-three DiFF $\widetilde{D}^{\sphericalangle,a}$ represented 
by the quark-gluon-quark correlator origins from the triparton correlation during the quark fragmentation \cite{Yang:2019aan}
\begin{eqnarray} \label{Eq:7} \nonumber
\widetilde{\Delta}_A^{\alpha}(z,k_T,R)&=&\frac{1}{2z} \sum_X \int\frac{d\xi^+d^2\xi_T}{(2\pi)^3}
e^{ik\cdot\xi}
\langle 0| \int_{\pm\infty^+}^{\xi^+} d\eta^+ {\cal{U}}^{\xi_T}_{(\infty^+,\xi^+)} \times g F_\bot^{-\alpha}{\cal{U}}^{\xi_T}_{(\eta^+,\xi^+)}\psi(\xi)|P_h,R;X\rangle  \\
&&\langle P_h,R;X|\bar{\psi}(0){\cal{U}}^{0_T}_{(0^+,\infty^+)}{\cal{U}}^{\infty^+}_{(0_T,\xi_T)}|0\rangle\mid_{\eta^+=\xi^+=0,\eta_T=\xi_T},
\end{eqnarray}
where $F_\perp^{-\alpha}$ is the gluon field strength tensor. After integrating out $\vec{k}_T$, one obtains \cite{Yang:2019aan}
\begin{eqnarray}\label{Eq:8}
\widetilde{\Delta}_A^\alpha(z,\cos\theta,M_h^2,\phi_R) = \frac{z^2 |\vec{R}|}{8M_h} \int d^2 k_T \tilde{\Delta}_A^\alpha(z,k_T,R).
\end{eqnarray}
By projecting out the usual Dirac structures, we obtain the following decomposition results
\begin{eqnarray} \label{Eq:9}
\widetilde{\Delta}_A^\alpha(z,\cos\theta,M_h^2,\phi_R) = \frac{R_T^\alpha}{16\pi z}\widetilde{D}^\sphericalangle \slashed{n}_-,
\end{eqnarray}
where the index $\alpha$ is restricted to be transverse. Thus $\widetilde{D}^\sphericalangle$ can be extracted from the correlator 
$\widetilde{\Delta}_A^\alpha(z,\cos\theta,M_h^2,\phi_R)$ by the following trace:
\begin{eqnarray} \label{Eq:10}
\frac{R_T^\alpha}{4\pi z} \widetilde{D}^\sphericalangle = \text{Tr}[\widetilde{\Delta}^\alpha(z,\cos\theta,M_h^2,\phi_R) \gamma^-],
\end{eqnarray}
where $\gamma^-$ is the negative light-cone Dirac matrix.

The DiFFs $D_1$ and $\widetilde{D}^\sphericalangle$ 
with flavor $a$ can be expanded in the relative partial waves of the dihadron system up to the $p$-wave level \cite{Bacchetta:2002ux}: 
\begin{eqnarray} \label{Eq:11.3}
D^a_{1}(z,\cos \theta,M_h^2)&=&D^a_{1,oo}(z,M_h^2)+D^a_{1,ol}(z,M_h^2)\cos \theta+D^a_{1,ll}(z,M_h^2)(3\cos^2\theta-1),\label{Eq:11.1} \\
\widetilde{D}^{\sphericalangle a}(z,\cos\theta,M_h^2)&=&\widetilde{D}^{\sphericalangle a}_{ot}(z,M_h^2)+\widetilde{D}^{\sphericalangle a}_{lt}(z,M_h^2)\cos\theta,
\end{eqnarray}
where $\widetilde{D}^\sphericalangle_{ot}$ comes from the interference of $s$- and $p$-waves, 
and $\widetilde{D}^\sphericalangle_{lt}$ originates from the interference of two $p$ waves with different polarization.

In this paper we will not consider the $\cos\theta$-dependent terms in the expansion of DiFFs. 
This is because that $\cos\theta$-dependent terms correspond to the higher order contribution in the partial wave expansion 
and can only be significant when the two hadrons produce via a spin-one resonance. 
Whereas the function $\widetilde{D}^{\sphericalangle}_{lt}$ can also contribute to a double spin asymmetry by integrating upon $\theta$ in a different range, 
$[ -\pi/2,\pi/2 ]$ and we will study this $\cos\theta$-dependent contribution in a future paper. 
Therefore, we focus on the functions $D_{1,oo}$, $H_{1,ot}^{\sphericalangle}$ and $\widetilde{D}^{\sphericalangle}_{ot}$. 
Under these selections, the $\cos\phi_R$ asymmetry of the considered process can be expressed as\cite{Sirtl:2017rhi}
\begin{eqnarray} \label{Eq:12} \nonumber
A_{LL}^{\cos\phi_R}&=&\frac{M}{Q}\frac{|R|}{M_h}\frac{\sum_a e_a^2 \left[ xe_L^a(x)H_{1,ot}^{\sphericalangle a}(z,M_h^2)-\frac{M_h}{Mz}g_1^a(x) \tilde{D}_{ot}^{\sphericalangle a}(z,M_h^2) \right]}{\sum_a e_a^2 f_1^a(x) D_{1,oo}^{a}(z,M_h)} \\
&=&-\frac{\sum_a e_a^2 \,\frac{|R|}{Qz}\, g_1^a(x) \tilde{D}_{ot}^{\sphericalangle a}(z,M_h^2)}{\sum_a e_a^2 f_1^a(x) D_{1,oo}^{a}(z,M_h)},
\end{eqnarray}
where we have applied the approximations of the diquark spectator model.

\section{The model calculation of $\widetilde{D}^\sphericalangle$}
\label{III}

Before working out the DiFF $\widetilde{D}^\sphericalangle_{ot}$ in the spectator model, 
we briefly review the calculation of twist-two DiFFs $D_{1,oo}$ given by Ref.\cite{Bacchetta:2006un}. 
The model can make predictions for these collinear DiFFs and also for transverse momentum-dependent DiFFs, 
which we will consider in a next work. As stated in Eq.(\ref{Eq:11.1}), $D_1$ 
was expanded in terms of the relative partial waves of the dihadron and the expansion was truncated up to the $p$-wave. 
The $D_{1,oo}$ can receive contributions from both $s$ and $p$ waves, but not from the interference between the two. 
The $s$- and $p$-wave quark dihadron vertex structures denoted by $F_s$ and $F_p$ are defined below 
and introduced in Ref.\cite{Bacchetta:2006un}. The vertex $F^p$ is complex. 
Finally, the parameters of the model are fixed by fitting the output of the PYTHIA Monte Carlo generator\cite{Sjostrand:2000wi}, 
and the numerical results of the twist-2 DiFFs $D_{1,oo}$ is therefore given.

\begin{figure}[htp]
\centering
\includegraphics[height=5.6cm,width=9.4cm]{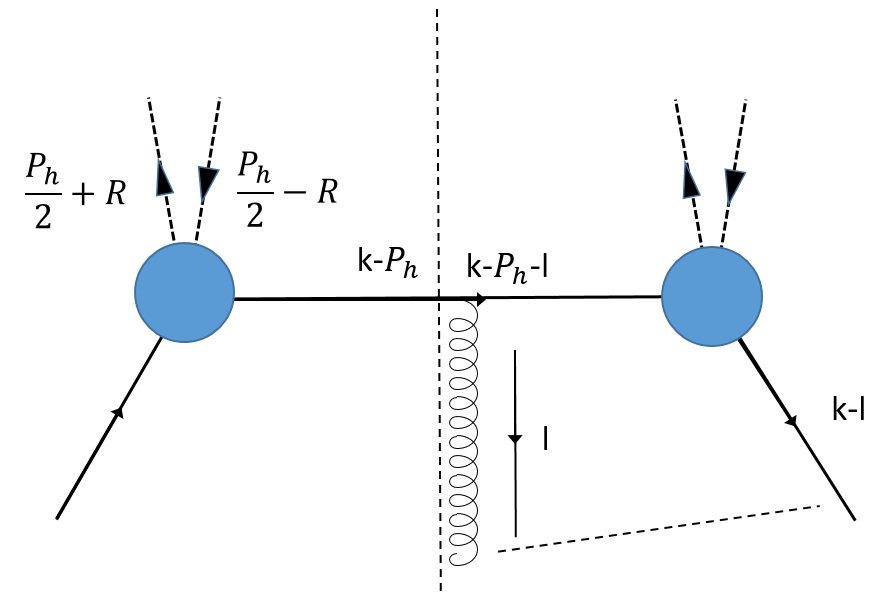}
\caption{ \normalsize
Diagrammatic representation of the correlation function $\widetilde{\Delta}_A^{\alpha}$ in the spectator model.}
\label{fig:2}
\end{figure}

We will work out the DiFF $\widetilde{D}^\sphericalangle$ in the spectator model in the following. 
It is desired to notice that we mainly following the calculation framework of Ref.\cite{Yang:2019aan}.
At the twist-three level, the DiFF $\widetilde{D}^\sphericalangle$ originates from the quark-gluon-quark correlator. 
The diagram adopted to calculate the twist-three DiFF $\widetilde{D}^\sphericalangle$ in the spectator model is shown in Fig.\ref{fig:2}. 
The left and right hand sides of Fig.\ref{fig:2} correspond to the quark-dihadron vertex $\langle P_h;X|\bar{\psi}(0)|0 \rangle$ 
and the vertex containing gluon rescattering $\langle 0|igF_\perp^{-\alpha}(\eta^+)\psi(\xi^+)|P_h;X \rangle$, respectively.
Here we adopt the Feynman gauge, in which the transverse gauge links ${\cal U}^{ \bm \xi_T}$ and ${\cal U}^{\bm 0_T}$ 
can be neglected \cite{Ji:2002aa,Belitsky:2002sm}.

With these considerations at hand, we can write down the s- and p-wave contributions to the quark-gluon-quark correlator as
\begin{eqnarray} \label{Eq:13}
\begin{aligned}
&\widetilde{\Delta}^\alpha (k,P_h,R)= i\frac{C_F\hat{\alpha}_s}{2{(2\pi)}^2(1-z)P_h^-}\frac{1}{k^2-m^2}\int 
\frac{d^4 \ell}{{(2\pi)}^4}(\ell^-g_T^{\alpha \mu}-\ell_T^\alpha g^{-\mu}) \\
&
\frac{(\slashed{k}-\slashed{\ell}+m)(F^{s*}e^{-\frac{k^2}{\Lambda_s^2}}+F^{p*}e^{-\frac{k^2}{\Lambda_p^2}}\slashed{R})
(\slashed{k}-\slashed{P}_h-\slashed{\ell}+M_s)\gamma_\mu (\slashed{k}-\slashed{P}_h+M_s)(F^{s}e^{-\frac{k^2}{\Lambda_s^2}}+
F^{p}e^{-\frac{k^2}{\Lambda_p^2}}\slashed{R})(\slashed{k}+m)}{(-\ell^-\pm i\varepsilon)((k-\ell)^2-m^2-i\varepsilon)((k-P_h-\ell)^2-M_s^2-i\varepsilon)(\ell^2-i\varepsilon)},
\end{aligned}
\end{eqnarray}
where $m$ and $M_s$ represent the masses of the fragmented quark and the spectator quark, respectively. 
$C_F$ represents the color factor $\frac{4}{3}$ and the strong coupling constant is denoted by $\hat{\alpha}_s$. 
The factor $\ell^- g_T^{\alpha \mu}-\ell_T^\alpha g^{-\mu}$ originates from the Feynman rule corresponding to the gluon field strength tensor, 
as described by the open circle in Fig.\ref{fig:2}. In Eq.(\ref{Eq:13}), in principle the Gaussian form factors should depend on the loop momentum $\ell$. 
Here following choice in Ref.\cite{Bacchetta:2007wc} we abandon this dependence and merely use $k^2$ instead of $(k-\ell)^2$ in those form factors 
to simplify the integration. This choice is introduced to cutting off the high-$k_T$ region. 
The s- and p-wave vertex structure $F^s$ and $F^p$ have the following forms \cite{Bacchetta:2006un}:
\begin{eqnarray} \label{Eq:14}
F^s&=&f_s \\
F^p&=&
f_\rho \frac{M_h^2-M_\rho^2-i\Gamma_\rho M_\rho}{M_h^2-M_\rho^2+\Gamma_\rho^2 M_\rho^2}+f_\omega 
\frac{M_h^2-M_\omega^2-i\Gamma_\omega M_\omega}{M_h^2-M_\omega^2+\Gamma_\omega^2 M_\omega^2} 
- i f_\omega'\frac{\sqrt{\lambda(M_\omega^2,M_h^2,m_\pi^2)}\Theta(M_\omega-m_\pi-M_h)}{4\pi \Gamma_\omega M_\omega^2 
[4M_\omega^2 m_\pi^2+\lambda(M_\omega^2,M_h^2,m_\pi^2)]^\frac{1}{4}},
\end{eqnarray}
where $\lambda(M_\omega^2,M_h^2,m_\pi^2)=[M_\omega^2-(M_h+m_\pi)^2][M_\omega^2-(M_h-m_\pi)^2]$ and $\Theta$ 
denotes the unit step function. The couplings $f_s$, $f_\rho$, $f_\omega$ and $f_\omega'$ are the model parameters. 
The first two terms of $F^p$ can be identified as the contributions of the $\rho$ and the $\omega$ resonances decaying into two pions. 
The masses and the widths of the two resonances can be obtained from the PDG \cite{Eidelman:2004wy}: $M_\rho=0.776$ GeV, 
$\Gamma_\rho=0.150$ GeV, $M_\omega=0.783$ GeV and $\Gamma_\omega=0.008$ GeV.
In addition, according to isospin symmetry we have the conclusion that the fragmentation correlators for processes 
$u \to \pi^+ \pi^- X$, $\bar{d} \to \pi^+ \pi^- X$, $d \to \pi^- \pi^+ X$ and $\bar{u} \to \pi^- \pi^+ X$ are the same.
Thus, by transforming the sign of $\vec{R}$, equivalently changing $\theta \to \pi-\theta$ and $\phi \to \phi+\pi$.
Therefore, the DiFF $\widetilde{D}^{\sphericalangle}$ which depends linearly on R coming from $d \to \pi^- \pi^+ X$ 
and $\bar{u} \to \pi^- \pi^+ X$ processes has an additional minus sign comparing to the $u \to \pi^+ \pi^- X$ process.

By using Eq.(\ref{Eq:8}) and expanding Eq.(\ref{Eq:13}), we obtain
\begin{eqnarray} \label{Eq:15}
\begin{aligned}
\widetilde{\Delta}^\alpha(z,\cos\theta,M_h^2,\phi_R)&=i\frac{C_F \hat{\alpha}_s z^2 |\vec{R}|}{32(2\pi)^5(1-z)M_h P_h^-} 
\int d|k_T|^2 \int d^4 \ell \frac{\ell^-g_T^{\alpha \mu}-\ell_T^\alpha g^{-\mu}}{k^2-m^2} \\
& \bigg[ |F_s|^2 e^{-\frac{2k^2}{\Lambda_s^2}} \frac{(\slashed{k}-\slashed{\ell}+m)(\slashed{k}-\slashed{P}_h
-\slashed{\ell}+M_s)\gamma_\mu (\slashed{k}-\slashed{P}_h+M_s)(\slashed{k}+m)}{(-\ell^-\pm i\varepsilon)((k-\ell)^2
-m^2-i\varepsilon)((k-P_h-\ell)^2-M_s^2-i\varepsilon)(\ell^2-i\varepsilon)} \\
&+ |F_p|^2 e^{-\frac{2k^2}{\Lambda_p^2}} \frac{(\slashed{k}-\slashed{\ell}+m)\slashed{R}(\slashed{k}-\slashed{P}_h
-\slashed{\ell}+M_s)\gamma_\mu (\slashed{k}-\slashed{P}_h+M_s)\slashed{R}(\slashed{k}+m)}{(-\ell^-\pm i\varepsilon)((k-\ell)^2-m^2
-i\varepsilon)((k-P_h-\ell)^2-M_s^2-i\varepsilon)(\ell^2-i\varepsilon)} \\
&+ F^{s*}F^p e^{-\frac{2k^2}{\Lambda_{sp}^2}} \frac{(\slashed{k}-\slashed{\ell}+m)(\slashed{k}-\slashed{P}_h
-\slashed{\ell}+M_s)\gamma_\mu (\slashed{k}-\slashed{P}_h+M_s)\slashed{R}(\slashed{k}+m)}{(-\ell^-\pm i\varepsilon)
((k-\ell)^2-m^2-i\varepsilon)((k-P_h-\ell)^2-M_s^2-i\varepsilon)(\ell^2-i\varepsilon)} \\
&+ F^s F^{p*} e^{-\frac{2k^2}{\Lambda_{sp}^2}} \frac{(\slashed{k}-\slashed{\ell}+m)\slashed{R}(\slashed{k}-\slashed{P}_h-
\slashed{\ell}+M_s)\gamma_\mu (\slashed{k}-\slashed{P}_h+M_s)(\slashed{k}+m)}{(-\ell^-\pm i\varepsilon)((k-\ell)^2-m^2-i\varepsilon)
((k-P_h-\ell)^2-M_s^2-i\varepsilon)(\ell^2-i\varepsilon)} \bigg].
\end{aligned}.
\end{eqnarray}
Here the $z$-dependent $\Lambda$-cutoffs $\Lambda_{sp}$ and $\Lambda_{s,p}$ have the relation
\begin{eqnarray}
\frac{2}{\Lambda_{sp}^2}=\frac{1}{\Lambda_s^2}+\frac{1}{\Lambda_p^2}.
\end{eqnarray}
where $\Lambda_{s,p}$ have the following ansatz :
\begin{eqnarray} \label{Eq:16}
\Lambda_{s,p}=\alpha_{s,p} z^{\beta_{s,p}}(1-z)^{\gamma_{s,p}}
\end{eqnarray}
and $\alpha$, $\beta$ and $\gamma$ are the parameters showing below. 
The $k^2$ term is fixed by the on-shell condition of the spectator
\begin{eqnarray} \label{Eq:17}
k^2=\frac{z}{1-z}k_T^2+\frac{M_s^2}{1-z}+\frac{M_h^2}{z} \ .
\end{eqnarray}
In Eq.(\ref{Eq:15}), the lines with $|F_s|^2$ and $|F_p|^2$ describe the pure $s$- and $p$-wave contributions, 
thus they will not make a difference in the interference of $s$- and $p$-waves functions $\widetilde{D}_{ot}^\sphericalangle$. 
While the last two lines describe the $s$- and $p$- wave interference, and they do contribute to the $\widetilde{D}_{ot}^\sphericalangle$.
Therefore, in principle there are two sources for nonvanish $\tilde{D}_{ot}^\sphericalangle$ at one loop level.
One is the real part of the loop integral over $\ell$, coupling with the real part of $(F^{s*}F^p+F^{s}F^{p*})$.
The other is the imaginary part of the loop integral over $\ell$, combined with the imaginary part of $(F^{s*}F^p+F^{s}F^{p*})$.
The real part of the integral is just the usual loop integral adopting the Feynman parameterization.
While for the imaginary part of the integral, we impose the Cutkosky cutting rules:
\begin{eqnarray} 
\begin{aligned}
\frac{1}{\ell^2+i\varepsilon} \to -2\pi i \delta(\ell^2) \qquad \qquad \frac{1}{(k-\ell)^2+i\varepsilon} \to -2\pi i \delta((k-\ell)^2).
\end{aligned}
\end{eqnarray}
Then we can obtain the final result for $\widetilde{D}_{ot}^\sphericalangle$,
\begin{eqnarray} \label{Eq:18}
\begin{aligned}
\widetilde{D}^\sphericalangle(z,M_h^2)&=\frac{C_F \hat{\alpha}_s z^2 |\vec{R}|}{8\pi^2(1-z)M_h} \int d|k_T|^2 \\
&\frac{1}{k^2-m^2} \bigg\{ M_s \Big[(k^2+m^2)A+(k^2-zk^2+zm^2-M_s^2+M_h^2)B\Big] \text{Im}[F^{s*}F^p] e^{-\frac{2k^2}{\Lambda_{sp}^2}}
\\
&+\frac{1}{4}m\Big[ (z-1)B_0(P_h,m,M_s)+(k^2-M_h^2+M_s^2)C_0(k^2,M_h^2,2k^2-M_s^2+2M_h^2,m,0,M_s)+
\\
&(k^2+M_h^2+M_s^2)C_1(k^2,2k^2-M_s^2+2M_h^2,M_h^2,m,0,M_s)-(k^2-M_h^2-M_s^2)
\\
&C_2(k^2,2k^2-M_s^2+2M_h^2,M_h^2,m,0,M_s) \Big] \text{Re}[F^{s*}F^p] e^{-\frac{2k^2}{\Lambda_{sp}^2}} \bigg\},
\end{aligned}
\end{eqnarray}
where $B_0,C_0,C_1$ and $C_2$ are the usual one loop scalar or tensor integrals.
The general defination of 2-point one-loop scalar integration is given by \cite{Ellis:2007qk} 
\begin{eqnarray}
\begin{aligned}
B_0(p_1^2;m_1^2,m_2^2)=\frac{1}{i\pi^2} \int d^4 l \frac{1}{(l^2-m_1^2+i\varepsilon)((l+q_1)^2-m_2^2+i\varepsilon)},
\end{aligned}
\end{eqnarray}
and three-point one-loop scalar integration is denoted as 
\begin{eqnarray}
\begin{aligned}
C_0(p_1^2,p_2^2,p_3^2;m_1^2,m_2^2,m_3^2)=\frac{1}{i\pi^2} \int d^4 \ell \frac{1}{(\ell^2-m_1^2+i\varepsilon)((\ell+q_1)^2-m_2^2+i\varepsilon)((\ell+q_2)^2-m_3^2+i\varepsilon)} ,
\end{aligned}
\end{eqnarray}
where $q_n \equiv \sum_{i=1}^n p_i$ and $q_0=0$. The coefficients $A$ and $B$ denote the following functions
\begin{eqnarray} \label{Eq:19}
A &=&\frac{I_1}{\lambda(k,M_h,M_s)}\left[ 2k^2(k^2-M_s^2-M_h^2)\frac{I_2}{\pi}+(k^2+M_h^2-M_s^2) \right], \\
B &=&-\frac{2k^2}{\lambda(k,M_h^2,M_s^2)}I_1\left[ 1+\frac{k^2+M_s^2-M_h^2}{\pi}I_2 \right],
\end{eqnarray}
which originate from the decomposition of the following integral \cite{Lu:2015wja}
\begin{eqnarray} \label{Eq:20}
\int d^4 \ell \frac{\ell^\mu \delta(\ell^2) \delta[(k-\ell)^2-m^2]}{(k-P_h-\ell)^2-M_s^2}=Ak^\mu + BP_h^\mu.
\end{eqnarray}
The functions $I_i$ represent the results of the following integrals
\begin{eqnarray} \label{Eq:21}
I_1&=&\int d^4 \ell \delta(\ell^2) \delta[(k-\ell)^2-m^2]=\frac{\pi}{2k^2}(k^2-m^2), \\
I_2&=&\int d^4 \ell \frac{\delta(\ell^2)\delta[(k-\ell)^2-m^2]}{(k-\ell-P_h)^2-M_s^2}=\frac{\pi}{2\sqrt{\lambda(k,M_h,M_s)}}\ln
\left( 1-\frac{2\sqrt{\lambda(k,M_h,M_s)}}{k^2-M_h^2+M_s^2+\sqrt{\lambda(k,M_h,M_s)}} \right).
\end{eqnarray}

\section{Numerical results}
\label{IV}

In order to fix the parameters of the spectator model, the authors of Ref.\cite{Bacchetta:2006un} 
compare it with the output of the PYTHIA event generator adopted for HERMES. 
The values of the parameters obtained by the fit are:
\begin{alignat}{3}
\alpha_s &=2.60\ \text{GeV} &\qquad& \beta_s=-0.751 &\qquad& \gamma_s=-0.193 \nonumber \\
\alpha_p &=7,07\ \text{GeV} &\qquad& \beta_p=-0.038 &\qquad& \gamma_p=-0.085 \\
f_s &=1197\ \text{GeV}^{-1} &\qquad& f_\rho=93.5 &\qquad& f_\omega=0.63 \nonumber\\
f_\omega' &=75.2 &\qquad& M_s=2.97M_h &\qquad& m=0.0\ \text{GeV}, \nonumber
\end{alignat}
where we have adopted the same choice as in Ref.\cite{Bacchetta:2006un} for the quark mass $m$ fixed to be zero GeV. 
Since the $\text{Re}(F^{s*}F^p)$ term is proportional to the quark mass $m$, 
in our calculation only the $\text{Im}(F^{s*}F^p)$ term in Eq.(\ref{Eq:18}) contributes to $\widetilde{D}^\sphericalangle$ numerically. 
Furthermore, we make a preliminary estimate for choosing the strong coupling $\hat{\alpha}_s \approx 0.3$.

\begin{figure}[htp]
\centering
\includegraphics[height=6.0cm,width=7.4cm]{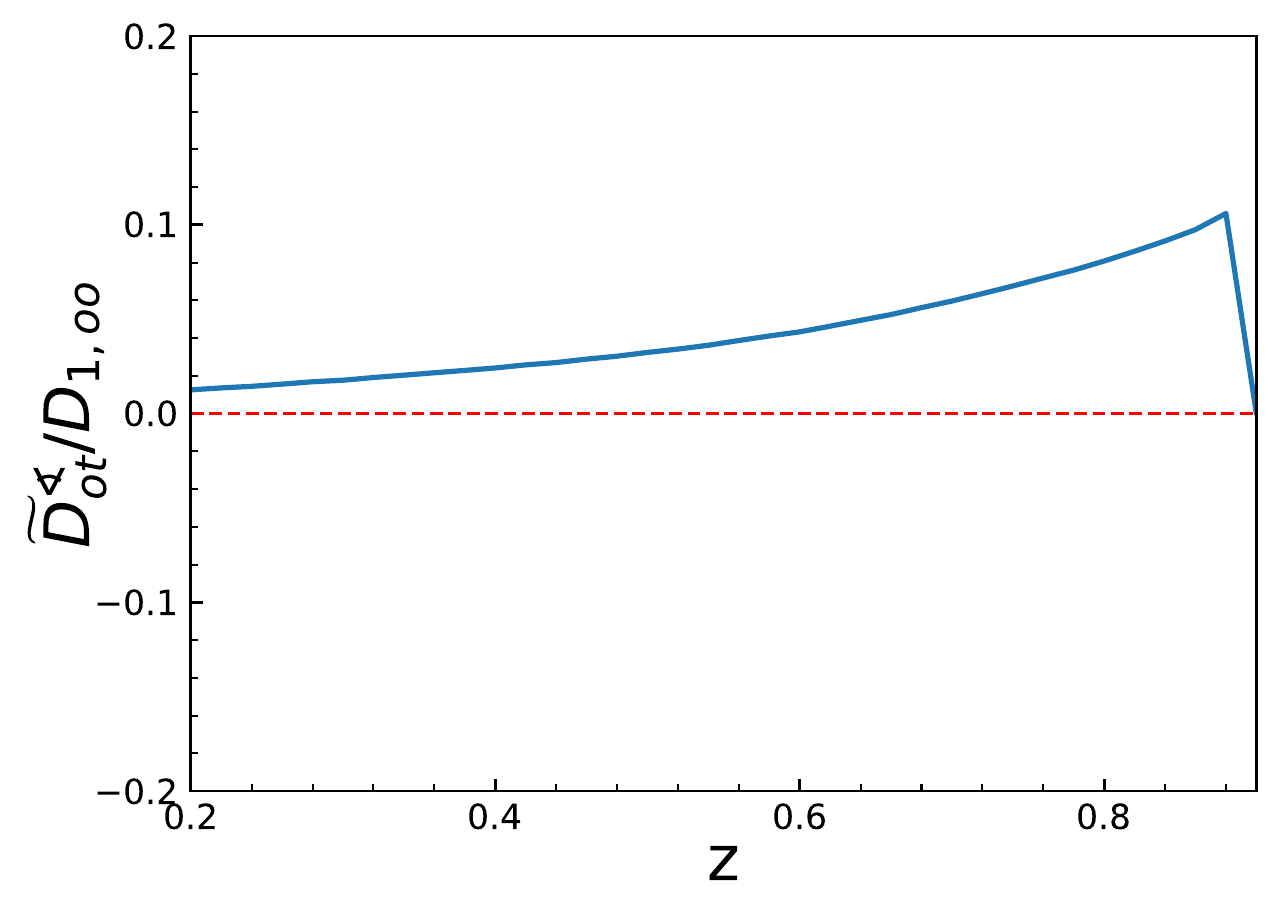}
\includegraphics[height=6.0cm,width=7.4cm]{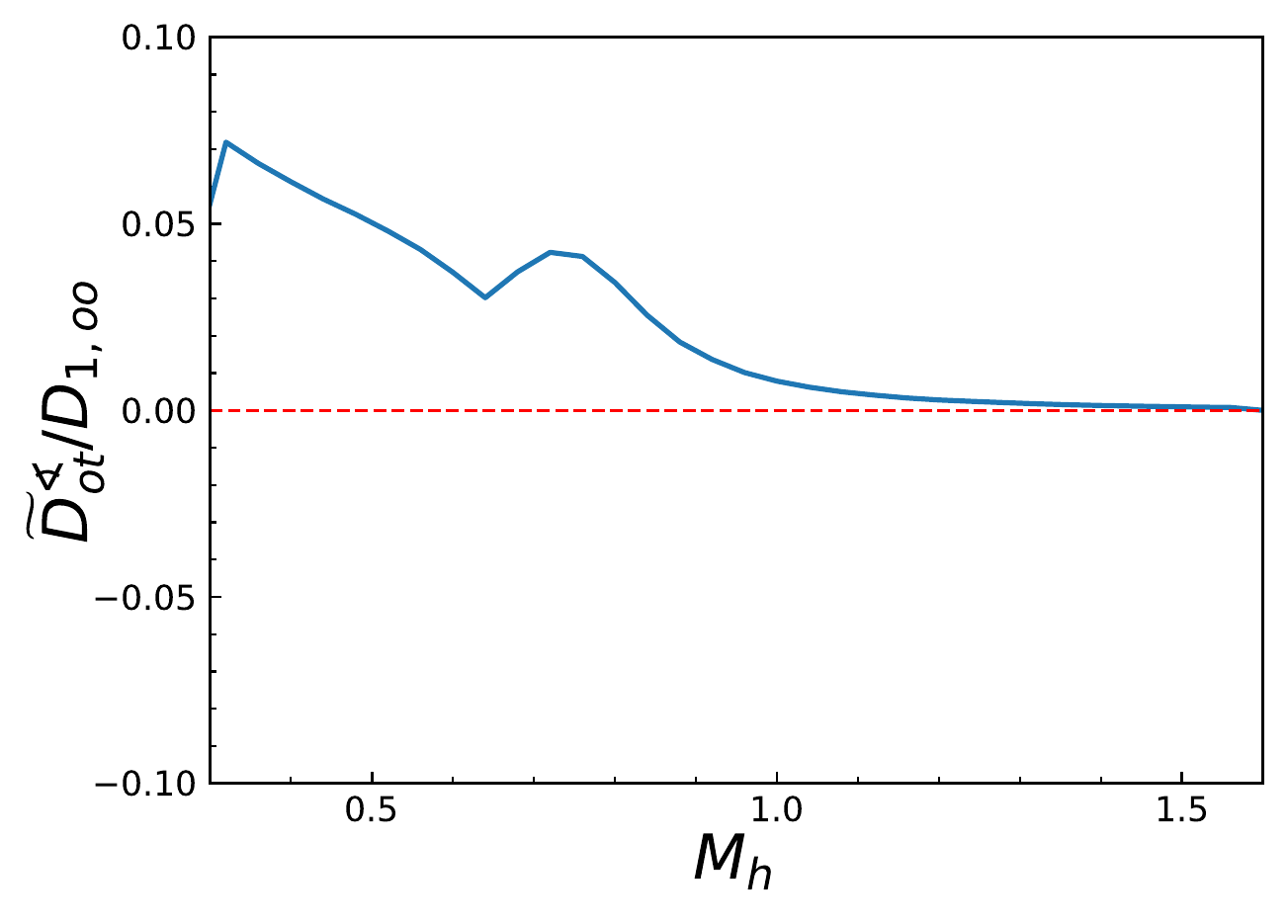}
\caption{\normalsize 
The twist-3 DiFF $\widetilde{D}^{\sphericalangle}_{ot}$ as functions of $z$ (left panel) 
and $M_h$ (right panel) in the spectator model, normalized  by the unpolarized DiFF $D_{1,oo}$.}
\label{fig:3}
\end{figure}
In the left panel of Fig.\ref{fig:3}, we plot the radio between $\widetilde{D}^\sphericalangle_{ot}$ and $D_{1,oo}$ as a function of $z$, 
integrated over the region $0.3\ \text{GeV}<M_h<1.6\ \text{GeV}$. In the right panel of Fig.\ref{fig:3} 
we plot the radio between $\tilde{D}^\sphericalangle_{ot}$ and $D_{1,oo}$ as a function of $M_h$ 
with $z$ integrated over the range $0.2<z<0.9$. Comparing with the unpolarized DiFF $D_{1,oo}$, 
the $\widetilde{D}^\sphericalangle_{ot}$ is an order of magnitude smaller.
The result in Fig.3 shows one example where a tilde-function is significantly smaller than a
non-tilde function, which provides a certain support for the Wandzura-Wilczek approximation and analog approximations as used for instance in Ref.\cite{Bastami:2018xqd}.

Then we present the numerical results of the $\cos\phi_R$ azimuthal asymmetry 
in the SIDIS process of longitudinally polarized muons off longitudinally polarized nucleon target. 
When expanding the flavor sum in the numerator of Eq.(\ref{Eq:12}), 
we apply the isospin symmetry mentioned in Sec.\ref{II} to the DiFF $\tilde{D}^\sphericalangle$. 
Furthermore, in principle sea quark distributions can be generated via perturbative QCD evolution and they are zero at the model scale. 
In this paper, we make a rough consideration by ignoring QCD evolution, which leads to zero antiquark PDFs $f_1$ and $g_1$.
Therefore, the expressions of the $x$-dependent, $z$-dependent and $M_h$-dependent $\cos\phi_R$ asymmetry can be adopted from Eq.(\ref{Eq:12}) as follows
\begin{eqnarray} \label{Eq:22}
A_{LL}^{\cos\phi_R}(x)&=&-\frac{\int dz \int dM_h 2M_h \frac{|\vec{R}|}{Qz}(4g_1^u(x)-g_1^d(x))
\tilde{D}^\sphericalangle_{ot}(z,M_h^2)}{\int dz \int dM_h 2M_h(4f_1^u(x)+f_1^d(x))D_{1,oo}(z,M_h^2)}, \\
\label{Eq:23}
A_{LL}^{\cos\phi_R}(z)&=&-\frac{\int dx \int dM_h 2M_h \frac{|\vec{R}|}{Qz}(4g_1^u(x)-g_1^d(x))
\tilde{D}^\sphericalangle_{ot}(z,M_h^2)}{\int dx \int dM_h 2M_h(4f_1^u(x)+f_1^d(x))D_{1,oo}(z,M_h^2)}, \\
\label{Eq:24}
A_{LL}^{\cos\phi_R}(M_h)&=&-\frac{\int dx \int dz \frac{|\vec{R}|}{Qz}(4g_1^u(x)-g_1^d(x))
\tilde{D}^\sphericalangle_{ot}(z,M_h^2)}{\int dx \int dz (4f_1^u(x)+f_1^d(x))D_{1,oo}(z,M_h^2)}.
\end{eqnarray}
We adopt the previous spectator model result for unpolarized DiFF $D_{1,oo}$ \cite{Bacchetta:2006un}. 
As for the twist-two PDFs $f_1$ and $g_1$, we apply the same spectator model results \cite{Bacchetta:2008af} for consistency.
To perform numerical calculation for the $\cos\phi_R$ asymmetry in dihadron SIDIS at COMPASS, we adopt the following kinematical cuts \cite{Sirtl}
\begin{eqnarray}
\begin{aligned}
&\sqrt{s} = 17.4\ \text{GeV} \qquad 0.003 < x < 0.4 \qquad 0.1 < y < 0.9 \qquad 0.2 < z < 0.9 \\
&0.3\text{GeV} < M_h < 1.6\ \text{GeV} \qquad Q^2 > 1\ \text{GeV}^2 \qquad W > 5\ \text{GeV},
\end{aligned}
\end{eqnarray}
where W is the invariant mass of photon-nucleon system with $W^2=(P+q)^2 \approx \frac{1-x}{x}Q^2$.
Since the evolution equations of the twist-3 DiFF $\tilde{D}^\sphericalangle$ is presently unknown, 
we disregard QCD evolution effects for all involved DIFFs and (for consistency) also for all PDFs, 
and use the model results at the low hadronic scale of the model. We present a first rough estimation for the asymmetry.
\begin{figure}[htp]
\centering
\includegraphics[height=6.0cm,width=5.7cm]{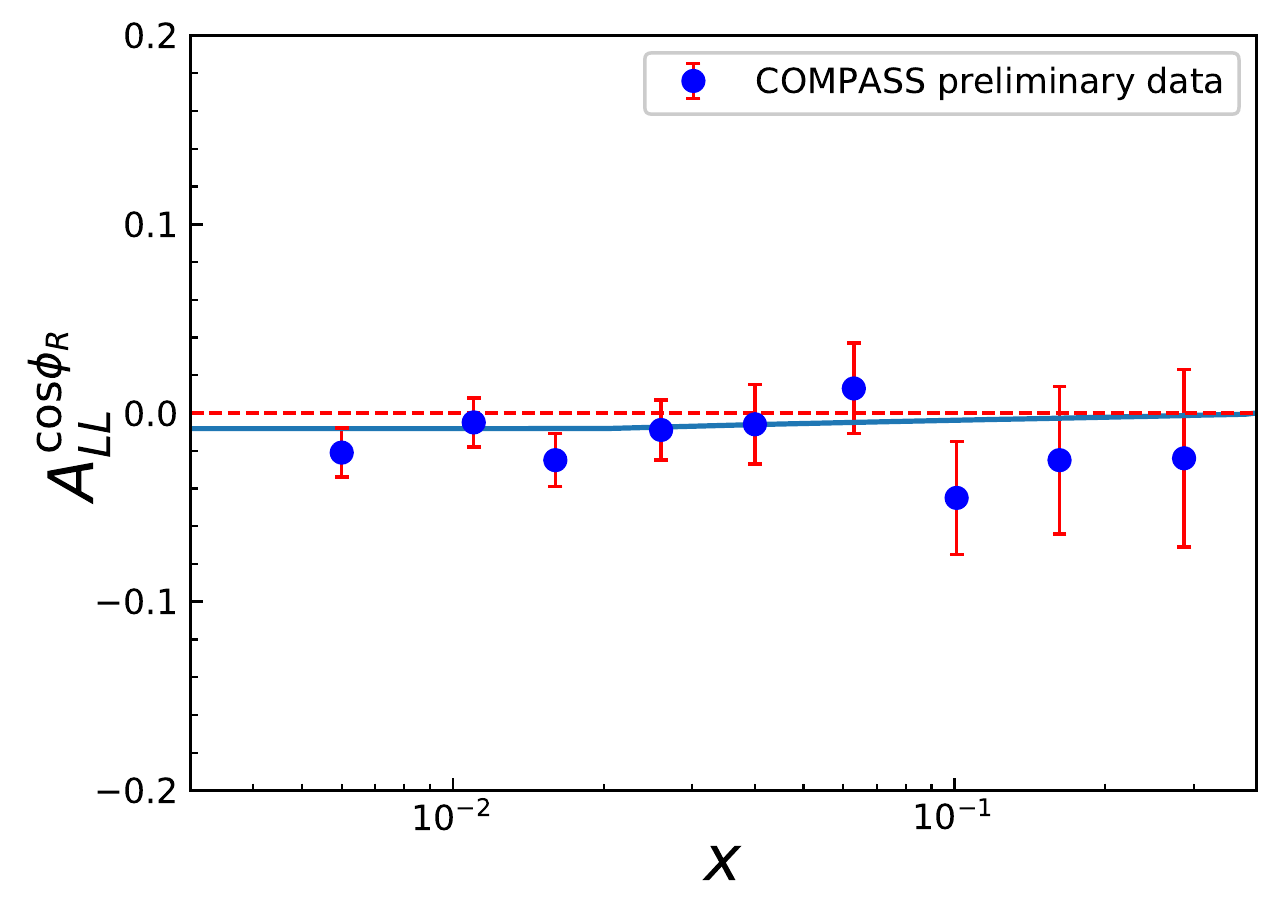}
\includegraphics[height=6.0cm,width=5.7cm]{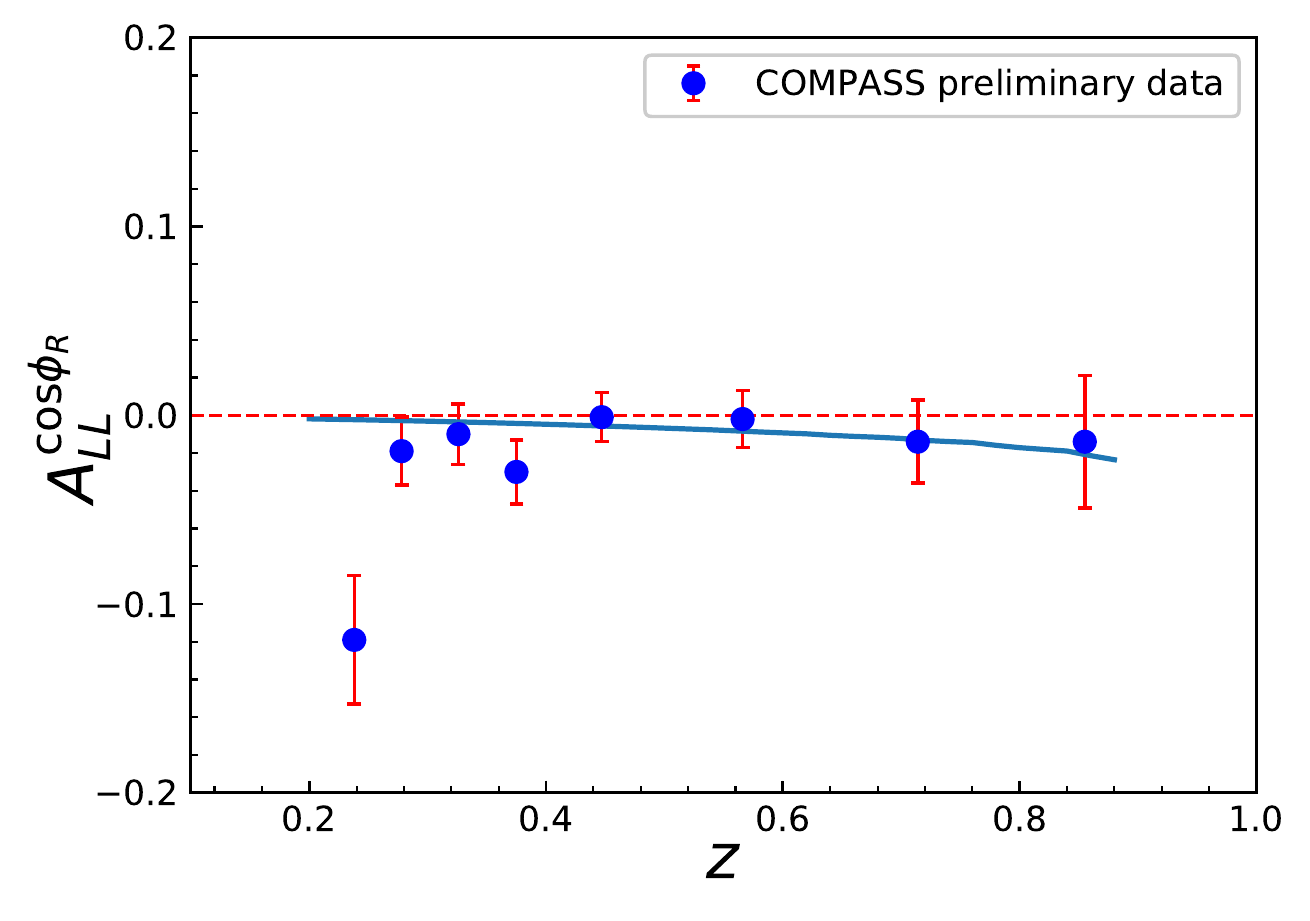}
\includegraphics[height=6.0cm,width=5.7cm]{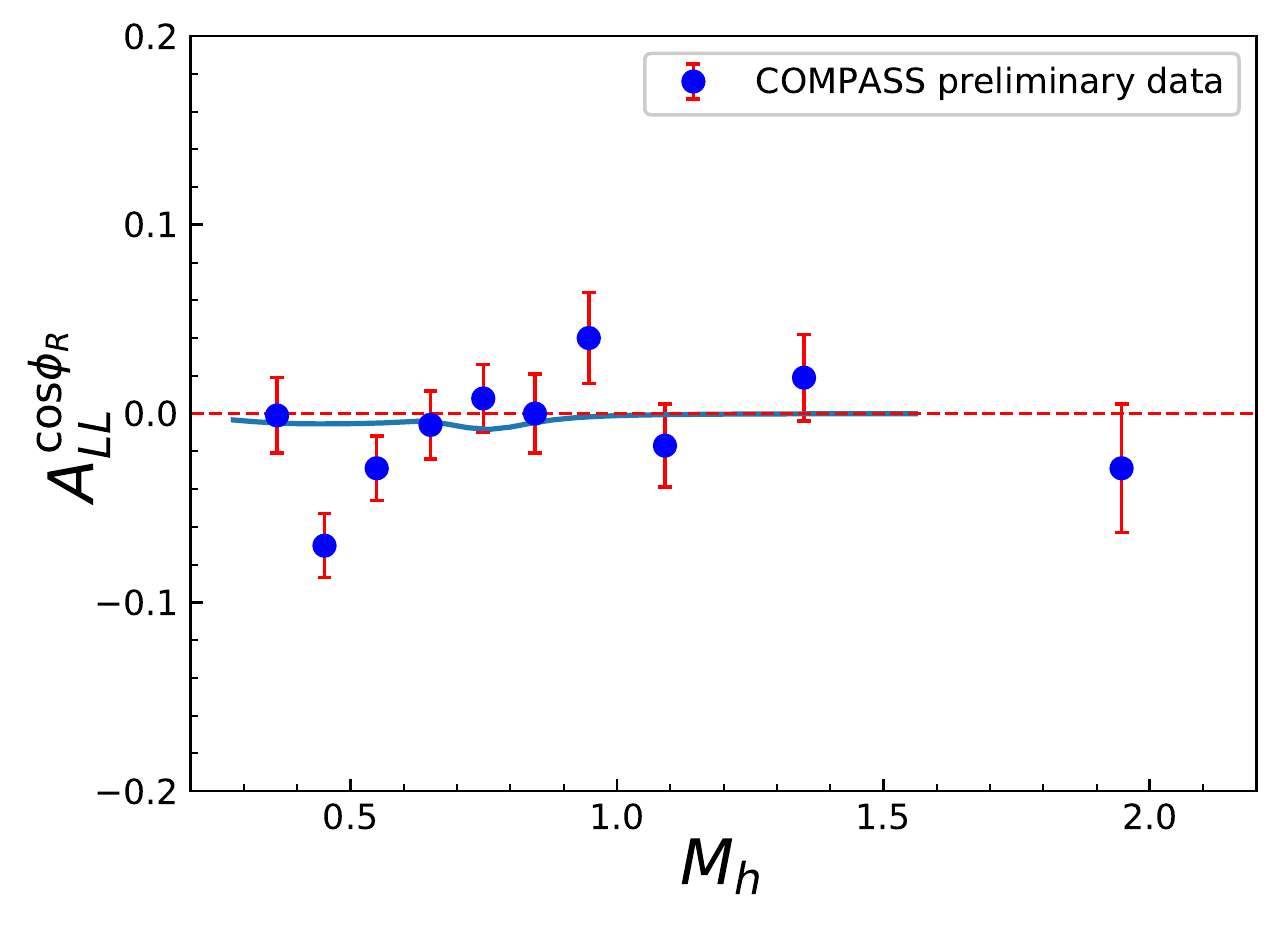}
\caption{\normalsize 
The $\cos\phi_R$ azimuthal asymmetry in the SIDIS process of longitudinally polarized muons off longitudinally polarized nucleon target 
as a functions of $x$ (left panel), $z$ (central panel) and $M_h$ (right panel) at COMPASS. 
The full circles with error bars show the preliminary COMPASS data for comparison. 
The solid curves denote the model prediction.}
\label{fig:4}
\end{figure}
Our main results in this work is our prediction for the $\cos\phi_R$ azimuthal asymmetry 
in the SIDIS process of longitudinally polarized muons off longitudinally polarized nucleon target, 
as shown in Fig.\ref{fig:4}. The $x$-, $z$- and $M_h$-dependent asymmetries are depicted in the left, 
central and right panels of the figure, respectively. The solid lines represent our model predictions. 
The full circles with error bars show the preliminary COMPASS data for comparison. 
Since we have not considered the QCD evolution effects, 
it can be found that the model results only give a rough prediction of the COMPASS preliminary data. 
In some probable future works, the obtained model results including the QCD evolution will give a reliable prediction.

\begin{figure}[htp]
\centering
\includegraphics[height=6cm,width=5.7cm]{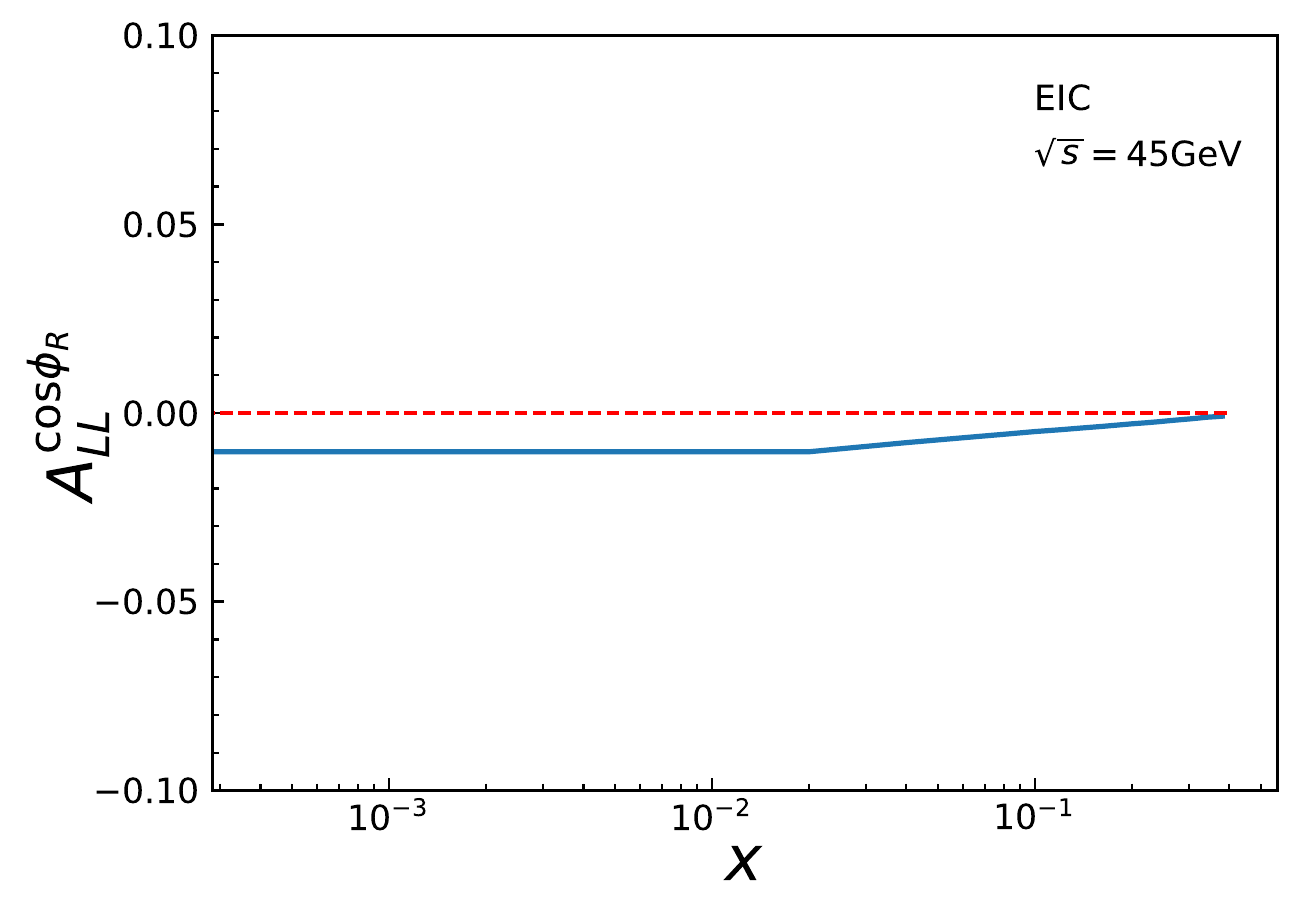}
\includegraphics[height=6cm,width=5.7cm]{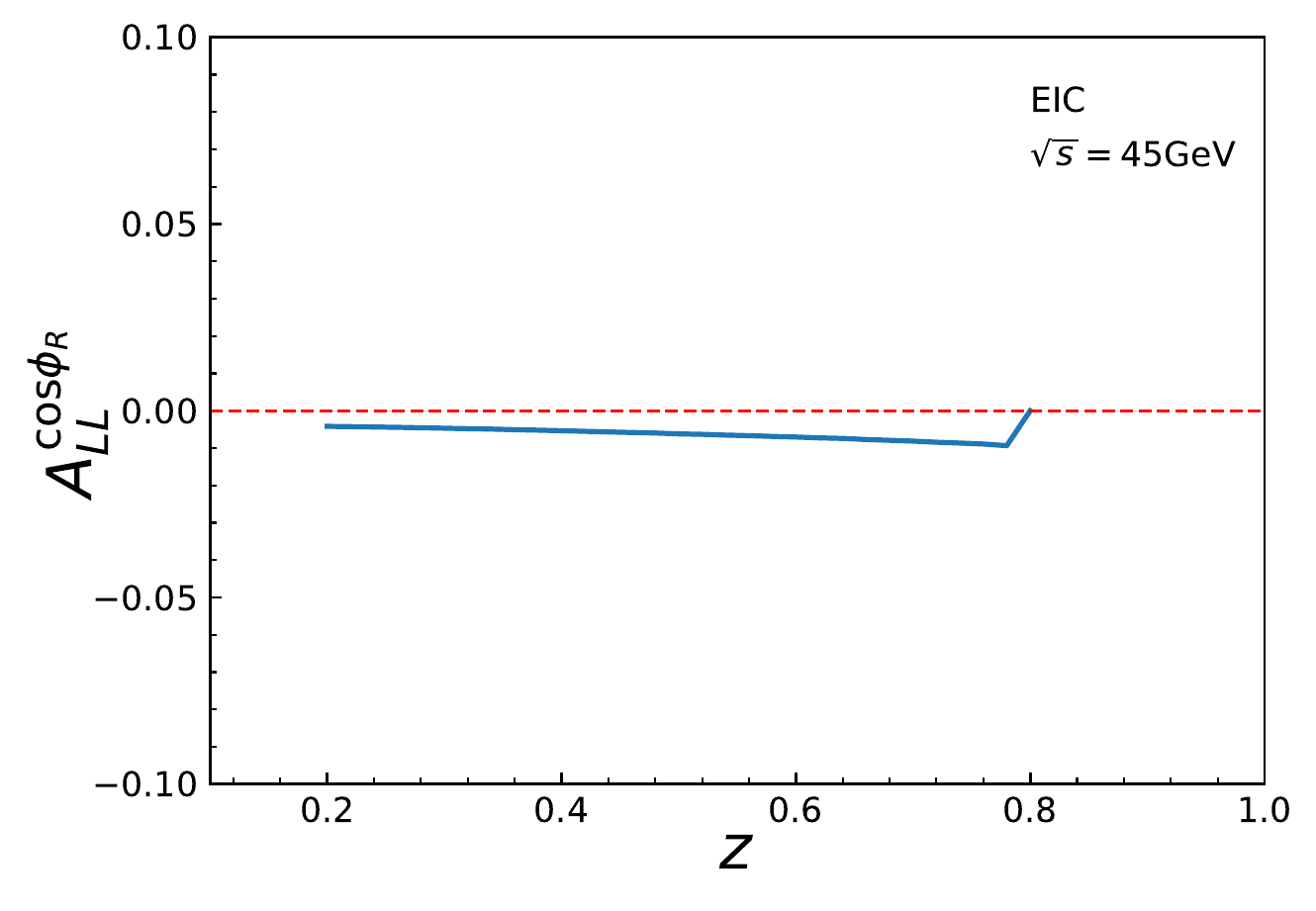}
\includegraphics[height=6cm,width=5.7cm]{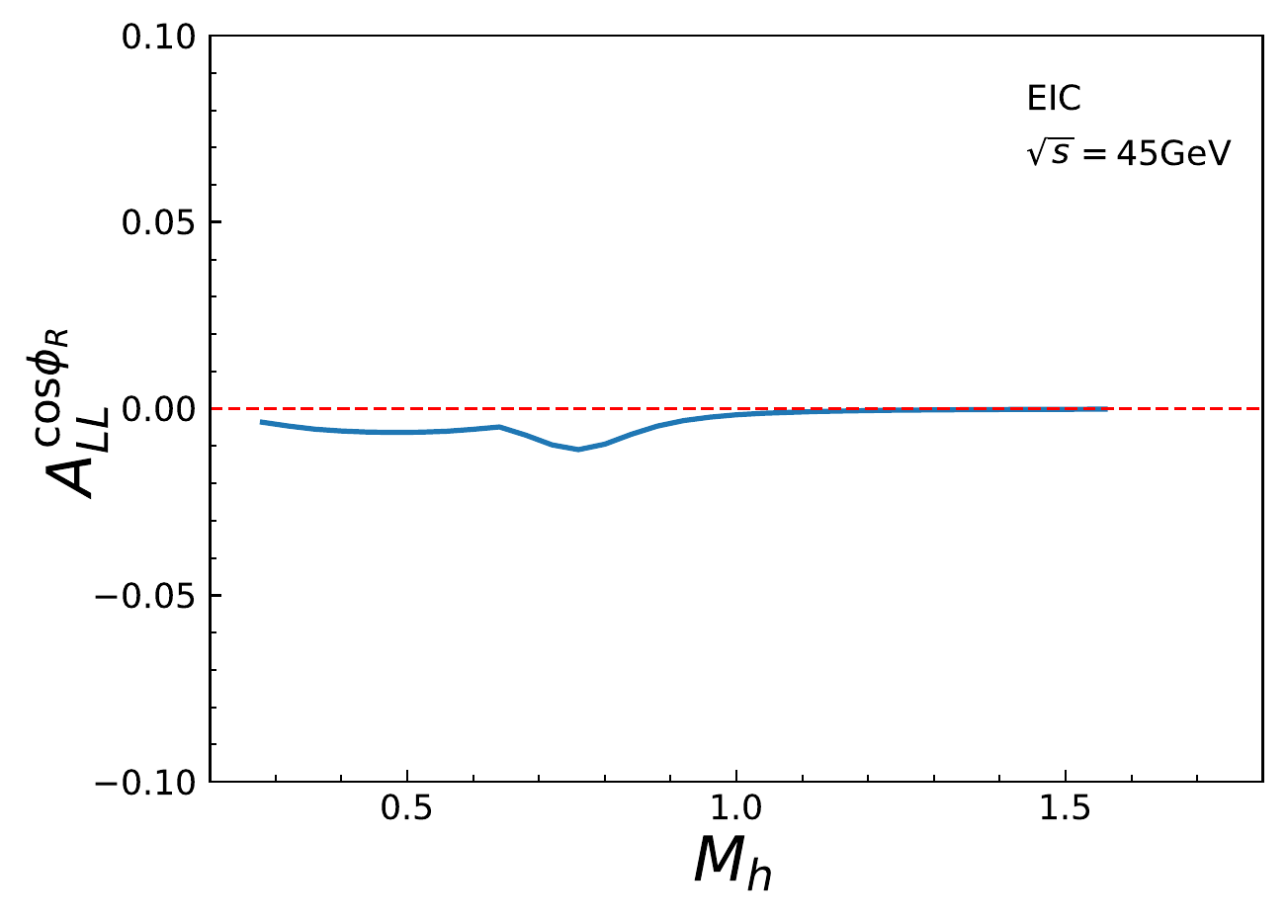}
\caption{\normalsize
The $\cos\phi_R$ azimuthal asymmetry in the SIDIS process of longitudinally polarized muons off longitudinally polarized nucleon target 
as a functions of $x$ (left panel), $z$ (central panel) and $M_h$(right panel) at the EIC. The solid curves denote the model prediction.}
\label{fig:5}
\end{figure}
In addition, to make a further comparison, we also make a prediction on the $\cos\phi_R$ asymmetry 
in the double longitudinally polarized SIDIS at the future EIC. Such a facility could be ideal to study this observable. 
We adopt the following EIC kinematical cuts \cite{Matevosyan:2015gwa}:
\begin{eqnarray}
\begin{aligned}
&\sqrt{s} = 45\ \text{GeV} \qquad 0.001 < x < 0.4 \qquad 0.01 < y < 0.95 \qquad 0.2 < z < 0.8 \\
&0.3\ \text{GeV} < M_h < 1.6\ \text{GeV} \qquad Q^2 > 1\ \text{GeV}^2 \qquad W > 5\ \text{GeV}.
\end{aligned}
\end{eqnarray}
The $x$-, $z$- and $M_h$-dependent asymmetries are plotted in the left, central, 
and right panels in Fig.\ref{fig:5}. We find that the overall tendency of the asymmetry at the EIC is similar to that at COMPASS. 
The size of the asymmetry is slightly smaller than that at COMPASS.

\section{Conclusion}
\label{V}

In this work, we have considered the double longitudinal spin asymmetry 
with a $\cos\phi_R$ modulation of dihadron production in SIDIS. 
With the spectator model result for $D_{1,oo}$ at hand, 
we worked out the twist-3 DiFF $\widetilde{D}^\sphericalangle_{ot}$ by considering the gluon rescattering effect. 
Using the partial wave expansion, we found that $\widetilde{D}^\sphericalangle_{ot}$ 
origins from the interference contribution of the $s$- and $p$-waves. 
By using the numerical results of the DiFFs and PDFs, 
we present the prediction for the $\cos\phi_R$ asymmetry and compare it with the COMPASS measurement. 
Since we have not considered the QCD evolution of PDFs and DiFFs, 
our result from $g_1 \widetilde{D}_{ot}^\sphericalangle$ coupling yields a preliminary and rough description of the COMPASS data. 
Moreover, we also estimate the $\cos\phi_R$ asymmetry in dihadron SIDIS at the typical kinematics of the future EIC. 
We conclude that in a spectator model calculation the twist-3 DiFF $\widetilde{D}^\sphericalangle_{ot}$ 
would be the dominate contribution so as to produce a preliminary understanding of the $\cos\phi_R$ asymmetry in dihadron production in SIDIS.

\begin{acknowledgments}
X. L. thanks professor Alessandro Bacchetta for his patient guidance.
The authors also thank professor Zhun Lu and doctor Yongliang Yang for useful discussions. 
H. S./ is supported by the National Natural Science Foundation of China (Grant No.11675033) 
and by the Fundamental Research Funds for the Central Universities (Grant No. DUT18LK27).
\end{acknowledgments}

\bibliography{v1}
\end{document}